\documentclass[sigconf,screen=true,bookmarks=false,natbib=true, 9pt]{acmart}
\usepackage{graphicx}
\usepackage{amsmath}
\usepackage{mathtools}
\usepackage{comment}
\usepackage[subrefformat=parens,labelformat=parens]{subfig}
\usepackage{bm}
\usepackage{multirow}
\usepackage{threeparttable,booktabs}
\usepackage{tikz}
\usepackage{balance}
\usepackage{courier}                                       
\usepackage{cleveref}                                      
\usepackage[mathcal]{eucal}
\usepackage[noend]{algpseudocode}
\algrenewcommand\textproc{\texttt}
\makeatletter\let\float@addtolists\relax\makeatother
\usepackage{algorithm}
\newcommand{\ForEach}[1]{\For{\textbf{each} #1}}
\usepackage{pgfplots}
\usepackage{pgfplotstable}
\pgfplotsset{compat=newest}
\usepackage{siunitx}
\let\norm\undefined 
\DeclarePairedDelimiter\norm{\lVert}{\rVert}

\theoremstyle{plain}

\theoremstyle{definition}

\newtheorem{myproblem}{\textbf{Problem}}

\usepackage[skip=1pt]{caption}            
\copyrightyear{2022}
\acmYear{2022}
\setcopyright{acmcopyright}\acmConference[ICCAD '22]{IEEE/ACM International Conference on Computer-Aided Design}{October 30-November 3, 2022}{San Diego, CA, USA}
\acmBooktitle{IEEE/ACM International Conference on Computer-Aided Design (ICCAD '22), October 30-November 3, 2022, San Diego, CA, USA}
\acmPrice{15.00}
\acmDOI{10.1145/3508352.3549384}
\acmISBN{978-1-4503-9217-4/22/10}

\begin{document}
\pagestyle{plain} 
\pagenumbering{gobble} 

\title{
TAG: Learning Circuit Spatial Embedding From Layouts
}

\author{
Keren~Zhu$^1$, Hao~Chen$^1$, Walker~J.~Turner$^2$, George~F.~Kokai$^2$,  Po-Hsuan~Wei$^2$,\\ David~Z.~Pan$^1$, and Haoxing~Ren$^2$}
\affiliation{
  \institution{$^1$ECE Department, The University of Texas at Austin, Austin, TX}
  \institution{$^2$NVIDIA Corporation}
  \institution{keren.zhu@utexas.edu, dpan@ece.utexas.edu, haoxingr@nvidia.com}
  \country{USA}
}

%
%
%
%
%
%
%

\begin{abstract}
Analog and mixed-signal~(AMS) circuit designs still rely on human design expertise.
Machine learning has been assisting circuit design automation by replacing human experience with artificial intelligence.
This paper presents TAG,  a new paradigm of learning the circuit representation from layouts leveraging \textbf{T}ext, self \textbf{A}ttention and \textbf{G}raph.
The embedding network model learns spatial information without manual labeling.
We introduce text embedding and a self-attention mechanism to AMS circuit learning.
Experimental results demonstrate the ability to predict layout distances between instances with industrial FinFET technology benchmarks.
The effectiveness of the circuit representation is verified by showing the transferability to three other learning tasks with limited data in the case studies: layout matching prediction, wirelength estimation, and net parasitic capacitance prediction.
\end{abstract}

\maketitle
\section{Introduction}
\label{sec:gackground}

    The performance of analog and mixed-signal~(AMS) integrated circuit designs are sensitive to parasitics, process variation, and layout-dependent effects.
    Today, AMS circuit design, from schematic to layout, is still mainly a manual, time-consuming, and error-prone task.

AMS circuits often impose specific parasitics and mismatch requirements on their layout implementation, where designers leverage their prior experience to place devices in specific patterns and configurations to reduce parasitics, the effects of local variation gradients, and layout-dependent effects.
Lacking the techniques to mimic such an experience automatically is one of the main bottlenecks in automating AMS design flow~\cite{MAGICAL_JOS20_Chen}.

Researchers have attempted to apply machine learning~(ML) to  AMS IC designs~\cite{Analog_JVLSI21_Afacan}.
Several studies use graph neural network~(GNN) on circuit graphs to learn the symmetry constraints in layouts~\cite{MAGICAL_DAC21_Chen, Analog_ASPDAC21_Gao}. 
The work~\cite{Analog_DATE20_Kunal} uses GNN to identify the type of AMS circuits, such as amplifiers and filters, to select layout templates for each circuit.
Researchers also represent schematics with graphs and use GNN for the analog device sizing problem~\cite{Analog_DATE20_Settaluri, Analog_DAC20_Wang}.
Wang et al.~\cite{CKTML_DAC22_Wang}  and Li et al.~\cite{CKTML_DAC22_Li} uses GNN to learn netlist representations based on their logic functionality.
The GNN-based ML frameworks decide transistors' width and length parameters based on the feedback from pre-layout simulations.
Netlists are essentially hyper-graphs, making GNN a solution candidate that many prior studies have adopted.
The underlying GNNs, in some sense, are expected to learn to capture the circuit representation.
Outside the AMS circuit domain, there are also attempts to learn the graph-structured circuit representation.
Wu et al.~\cite{Other_DAC22_Wu} investigated learning on high-level synthesis codes with GNN.
Several studies apply ML to learn source code representation~\cite{ML_PAPL19_Alon, ML_ICLR20_Hellendoorn}.
However, despite its wide adoption, circuit representation learning is seldom studied as an individual problem.
The underlying neural network models are trained with different targets in the individual applications. 
In this work, we propose a new paradigm to learn AMS circuit representation without additional manual labeling by leveraging the layout/placement data directly.

Circuits are commonly represented as graphs, and existing learning algorithms apply GNN on the circuit graph.
Nonetheless, the knowledge of the graph representation is limited to device connectivity only.
However, detailed information is readily available in the circuit netlist in the form of the device, instance, and net names, where designers often use specific naming conventions to detail and organize the netlist to be more human readable.
To leverage this information, We adopt the natural language model in the representation learning process. 
On the other hand, the graph convolution mechanism is usually limited in capturing a global view of the entire circuits. 
To address this, we also adopt a sub-circuit-wise self-attention mechanism to integrate the whole picture of the sub-circuit into the resulting embedding.

\begin{figure}[t]
    \centering
    \includegraphics[width=0.5\textwidth]{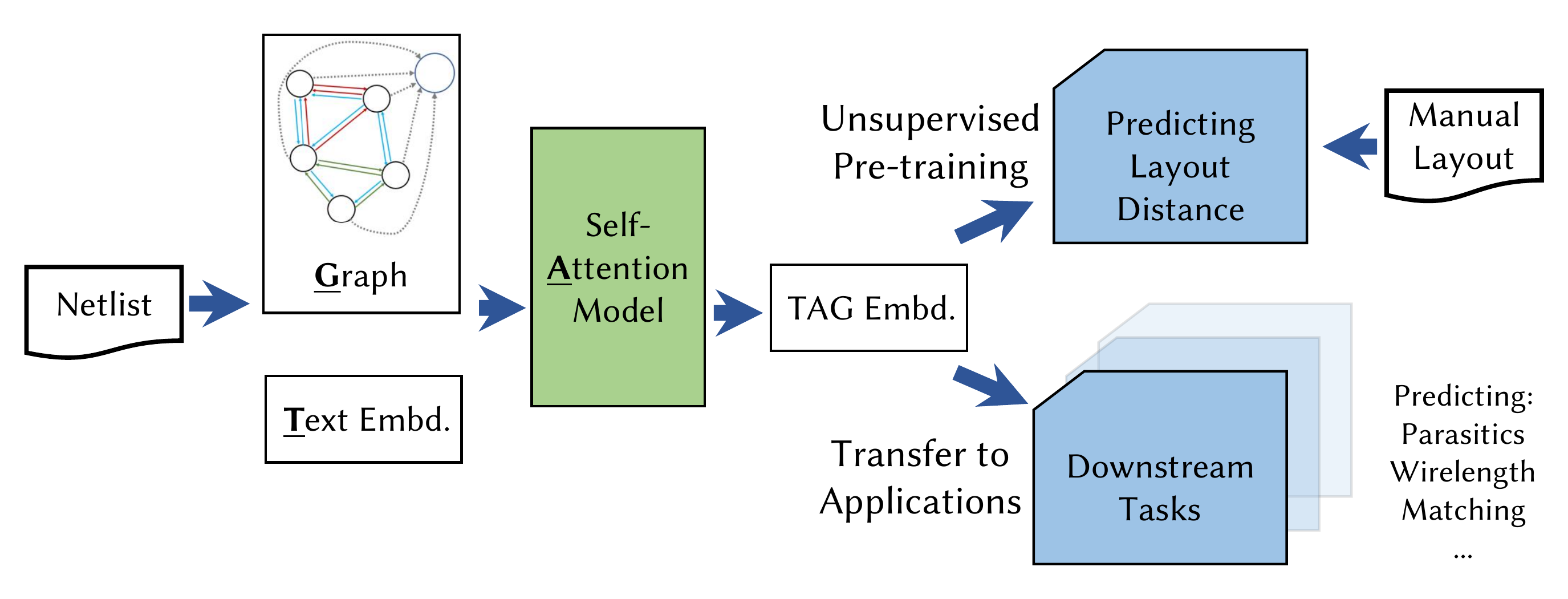}
    \caption{An illustration of the proposed circuit representation learning paradigm.}
    \label{fig:paradigm}
\end{figure}

In this paper, we propose TAG (\textbf{T}ext, Self-\textbf{A}ttention and \textbf{G}raph), a framework that learns the AMS circuit representation from layout positions.
Inspired by the success of the pre-trained model in natural language models (e.g., BERT~\cite{ML_ARXIV19_Devlin}), the TAG framework pre-trains a learning model on a larger layout dataset first.
Then the pre-trained circuit instance embeddings can be used for other learning tasks with limited data.
Figure~\ref{fig:paradigm} shows our proposed circuit learning paradigm. 
A design database of netlists and corresponding manual layouts are used to pre-train a TAG model.
The pre-trained TAG models are then transferable to multiple applications in analog CAD.
This work attempts to establish a common learning representation for analog circuits.
The main contribution of this work is summarized as follows.
\begin{itemize}
    \item A framework, TAG, to learn and embed the circuit instance representation from layout data without additional manual labeling is presented. It learns the spatial relations of instances in the embedding and assists in transferring to other learning tasks.
    \item A novel methodology of incorporating the circuit netlist text information, such as the instance and type names, from netlists into the learning task is proposed. 
    \item A circuit embedding network combining a multi-head self-attention layer with GNN is presented. The proposed usage of the self-attention mechanism allows the resulting instance embedding to reflect a better global view of the circuits.
    \item Experimental results show TAG significantly outperforms the existing methods in the accuracy of predicting relative layout distance. TAG also demonstrates great effectiveness in transferring to three other learning tasks: layout matching prediction, wirelength estimation, and net parasitic capacitance prediction.
\end{itemize}

The remainder of this paper is organized as follows.
Section~\ref{sec:prelim} gives the preliminaries.
Section~\ref{sec:algo} details the proposed TAG framework.
Section~\ref{sec:result} presents the experimental results, and Section~\ref{sec:concl} concludes the paper.
\section{Preliminaries} \label{sec:prelim}

In this section, we introduce the convolutional graph neural network~(Section~\ref{sec:prelim:gnn}).
Then we describe the circuit hierarchical structure and formulate our learning target: the relative instance distance~(Section~\ref{sec:prelim:problem}).
In the end, we overview three applications that are used for case studies in the experiments~(Section~\ref{sec:prelim:case}).


\subsection{Convolutional Graph Neural Network}
\label{sec:prelim:gnn}

Convolutional graph neural networks~(ConvGNN) are widely used for graph-structured data.
ConvGNNs perform \textit{convolution} on graph structures to obtain new node embeddings.
For a node $n_i$, a graph convolution operation aggregates the current embedding or feature of $n_i$'s neighboring nodes as shown in the Equation~\eqref{eq:convgnn},
\begin{equation}
    \label{eq:convgnn}
    \begin{aligned}
    \mathbf{a}^{l+1}_i &= \mathrm{AGGREGATE}^l\left(\left\{ \mathbf{h}_j^{l}  : u \in \mathcal{N}_i\right\}\right), \\
    \mathbf{h}^{l+1}_i &= \mathrm{COMBINE}^l(\mathbf{h}_i^{l}, \mathbf{a}_i^{l+1}),
    \end{aligned}
\end{equation}
where $\mathbf{h}^{l}_i$ is the $l^\mathrm{th}$ layer output embedding for $n_i$, $\mathcal{N}_i$ indicates the neighbors of node $n_i$.
The choice of $\mathrm{AGGREGATE}(\cdot)$ and  $\mathrm{COMBINE}(\cdot)$ functions vary in ConvGNN layer designs.
A typical practice is to use pooling functions, such as $\mathrm{max}$ and $\mathrm{mean}$, and linear transformations.
After passing through the graph convolution layers, the node embedding can be used for some downstream prediction tasks.

While ConvGNNs have demonstrated success in many applications, there are several limitations in representative ConvGNN architectures.
First, the knowledge learned from ConvGNNs tends to have a strong locality.
ConvGNNs usually work by aggregating neighboring information, which results in similar behavior to low-pass filters on the graph spectral domain~\cite{ML_ICML19_Wu}.
Therefore, ConvGNNs sometimes may lack a global view of the graph.
Second, ConvGNNs can hardly distinguish locally isomorphic structures.
It becomes an issue as AMS designs frequently contain symmetric or parallel local structures.

\subsection{Sub-Circuits and Relative Layout Distance}
\label{sec:prelim:problem}

AMS circuit designs are intrinsically hierarchical.
Both the schematic and the layout design are usually implemented hierarchically.
A sub-circuit, such as a current mirror and an OTA, can function individually and be used as the building blocks for different top-level circuits.
On the other hand, normalizing the learning task to the sub-circuit scale allows a more general inductive basis in the learning model.
It benefits the transferability of the ML model to circuits with different scales. 
Therefore we focus on the sub-circuit-level in our unsupervised training scheme.

\begin{figure}[t]
    \centering
    \includegraphics[width=0.22\textwidth]{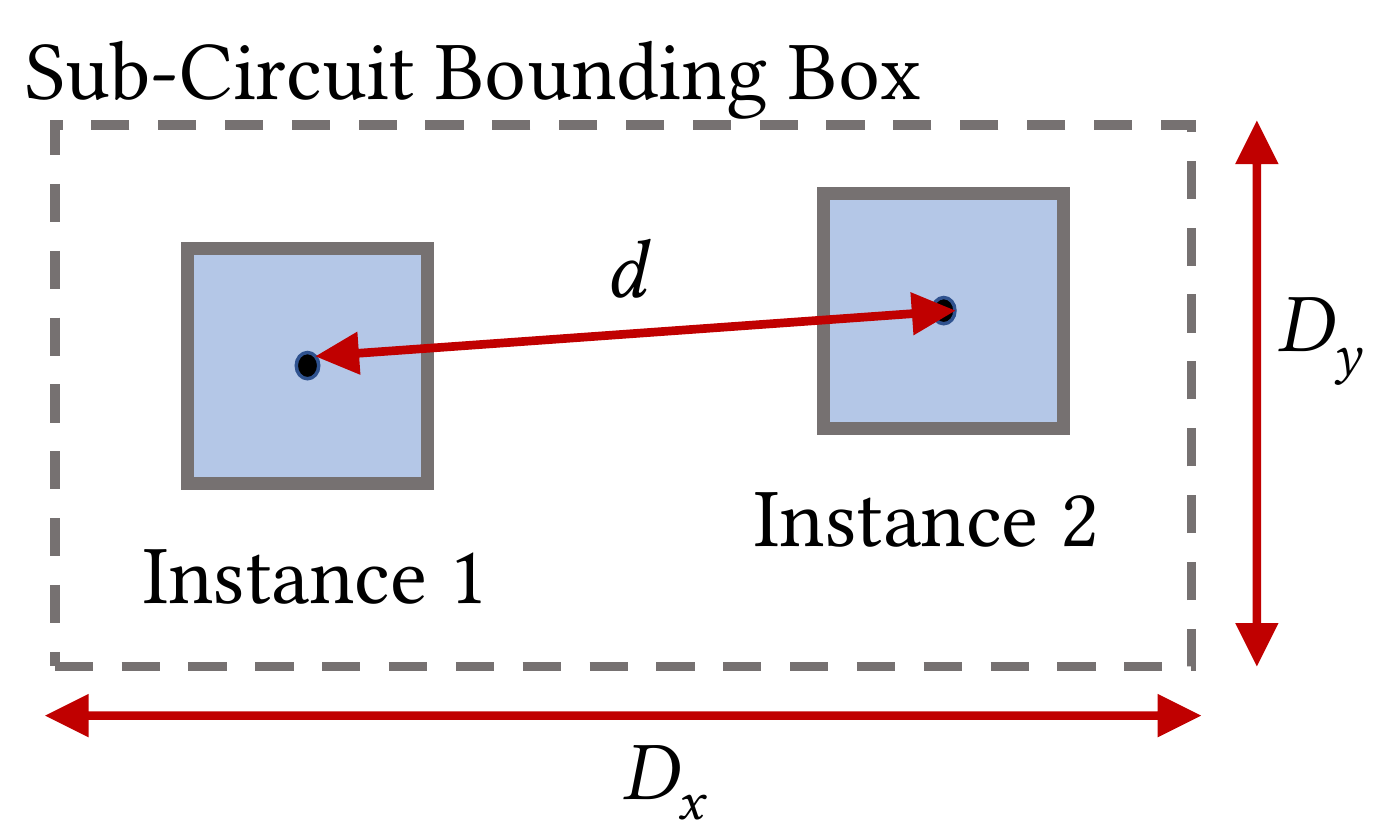}
    \caption{An example of the layout distance.}
    \label{fig:relative_dist}
\end{figure}

AMS layout requires careful considerations of parasitics, matching, area, power, etc. 
Specifically, layout constraints are commonly employed to ensure proper matching, device interdigitation, symmetry placement, and distances to critical signal paths.
We propose learning distance between instances as it is a crucial measurement from layout implementation and containing the design expertise.
Depending on the hierarchy level and the design, the instances can be either a primitive device~(e.g., transistor) or a sub-circuit~(e.g., OTA).
To allow the learning model to work for arbitrary circuits, we normalize the distance by its parent circuit bounding box so that its value is between 0 and 1.
Figure~\ref{fig:relative_dist} shows an example of the layout distance.
The distance $d$ between every instance pair is normalized to $\hat{d} = d / \sqrt{D_x^2 + D_y^2}$, where $D_x$ and $D_y$ denotes the width and height of its parent sub-circuit layout bounding box.
The normalized distance is used as our training target.
Such practice also motivates our ML model to homogeneously learn the knowledge between different hierarchy levels.

The relative layout distance prediction learning task is formulated as follows.
\begin{myproblem}[Relative Layout Distance Prediction]
Given a circuit design $D$ with hierarchy tree structure with a set of sub-circuits $C$,
predict the relative distance $\hat{d}$ in the manual layout implementations between all instance pairs ($\{I_i, I_j\} \in I$) in the same sub-circuit $C_i \in C$.
\end{myproblem}

\subsection{Applications in Analog CAD}\label{sec:prelim:case}

In this paper, we introduce three downstream applications as case studies to evaluate our proposed circuit embedding.

\subsubsection{Layout Matching Detection}
Identifying matching constraints in sub-circuits is crucial for fully-automated layout syntheses~\cite{MAGICAL_ASPDAC22_Zhu_b}. 
The matched instances are placed in certain matching patterns, such as symmetry and common-centroid.
The identification problem can be formulated as a binary classification problem, for which GNN is recently leveraged to solve~\cite{Analog_ICCAD20_Kunal, MAGICAL_DAC21_Chen, Analog_ASPDAC21_Gao}.

In the case study, we formulate the layout matching detection problem as follows.
\begin{myproblem}[Layout Matching Detection]
Given a circuit design  $D$ with a hierarchy tree structure with a set of sub-circuits $C$,
for every instance pairs ($\{I_i, I_j\} \in I$) that in the same sub-circuit $C_i \in C$, predict whether it is forming a symmetry, common centroid, or interdigitation patterns in the human layout implementation.
\end{myproblem}

\subsubsection{Wirelength Estimation}

A priori wirelength estimation is a classical problem in VLSI design automation~\cite{PD_ICCAD05_Kahng}.
It guides the early design stages.
Modern algorithms leverage ML techniques to increase the accuracy of the estimator~\cite{PD_DATE19_Hyun}.

In the current analog layout synthesis framework, the weights of nets and the proximity of instances are usually treated as human-specified parameters.
Finding a suitable set of parameters requires design expertise and trial and error~\cite{MAGICAL_ASPDAC20_Liu}.
A wirelength estimator can assist this process.

In the case study, we formulate the wirelength estimation problem as follows.
\begin{myproblem}[Wirelength Estimation]
Given a circuit design $D$ with a hierarchy tree structure with a set of sub-circuits $C$,
for every net $n_i$ in the same sub-circuit $C_i \in C$, predict its half-perimeter wirelength~(HPWL) in the human layout implementation.
\end{myproblem}

\subsubsection{Net Parasitic Capacitance Prediction}\label{sec:prelim:case:para}

Predicting post-layout parasitics from the schematic is an important problem in advanced technology nodes where the mismatch of pre-layout simulation and post-layout performance is significant.
Researchers have introduced ML methods to tackle the problem~\cite{PD_ML_DAC20_Shook, PD_ML_DAC20_Ren}.
By predicting the post-layout parasitics from schematics, those methods reduce the error of pre-layout simulation and accelerate the design cycle.

The state-of-the-art algorithm, ParaGraph~\cite{PD_ML_DAC20_Ren}, introduces GNN to the problem.
For each parasitic type, such as net capacitance, it trains multiple models for a different range of values.
Each model is specified with a maximum prediction value~($max_v$).
The models are then merged using the ensemble modeling technique to produce the final prediction.
Such methodology benefits the overall accuracy by allowing the models to focus on a small range of magnitude of regression targets.

We apply our TAG embedding in our experiments to the most representative parasitics prediction task: the net capacitance prediction problem.
We formulate the problem as follows.
\begin{myproblem}[Net Parasitic Capacitance Prediction]
Given a circuit design $D$ with a hierarchy tree structure with a set of sub-circuits $C$,
for every net $n_i$  in the flatten netlist $\hat{C}$, predict its post-layout total parasitic capacitance in human layout implementation.
\end{myproblem}

\section{TAG Algorithms}
\label{sec:algo}

TAG's circuit embedding network architecture comprises a GNN and a multi-head self-attention layer~(MSA).
The GNN model works on the entire hierarchical circuit to obtain the initial embeddings.
To mitigate the locality of the GNN model, we use the MSA layer on the sub-circuit instances to allow the resulting embeddings to consider the entire sub-circuit.
We add instance text embeddings during the MSA step to provide an additional dimension of knowledge.
We train the embeddings by regressing to relative layout distance.

In the rest of this section, we present the details of the algorithms.
The graph structure for GNN learning is shown in Section~\ref{sec:algo:graph}.
The instance input features are described in Section~\ref{sec:algo:feature}.
The embedding network architecture is presented in Section~\ref{sec:algo:embed}.
Finally, we introduce the learning algorithm for relative layout distance regression in Section~\ref{sec:algo:dist}.

\subsection{Heterogeneous Hierarchical Graph Construction}
\label{sec:algo:graph}

We propose to use a heterogeneous hierarchical Graph $G = (V, E)$ to represent a circuit.
At the device level, we adopt a similar approach to the work~\cite{MAGICAL_DAC21_Chen}.
Each device is represented as a node, and the nets are decomposed into two-pin pairs.
A directed edge $e = (u, v, \tau_v) \in E$ indicates the interconnection from vertex $u$ to $v$ with edge type $\tau_v$.
The edge type denotes the type of the connected port of $v$ in $e$.
The port type can be the gate, drain, source, passive device, and sub-circuit.
The power and ground nets are not extracted into the graph as they trivialize the graph by connecting most of the nodes.
We exclude the dummy and decap devices in the graph and learning process as they are mainly used to compensate for layout effects instead of functioning in circuits.

Different from the work~\cite{MAGICAL_DAC21_Chen}, the circuit hierarchy is incorporated in the graph.
Each sub-circuit is also represented as a node in the graph.
A directed edge $e = (u, v, \tau_{hier})$ is added from the child node $u$ to its parent $\mathrm{parent}(v)$.
The backward parent-children edges are not added with the assumption that the circuit implementations are bottom-up.

\begin{figure}[t]
    \centering
    \includegraphics[width=0.5\textwidth]{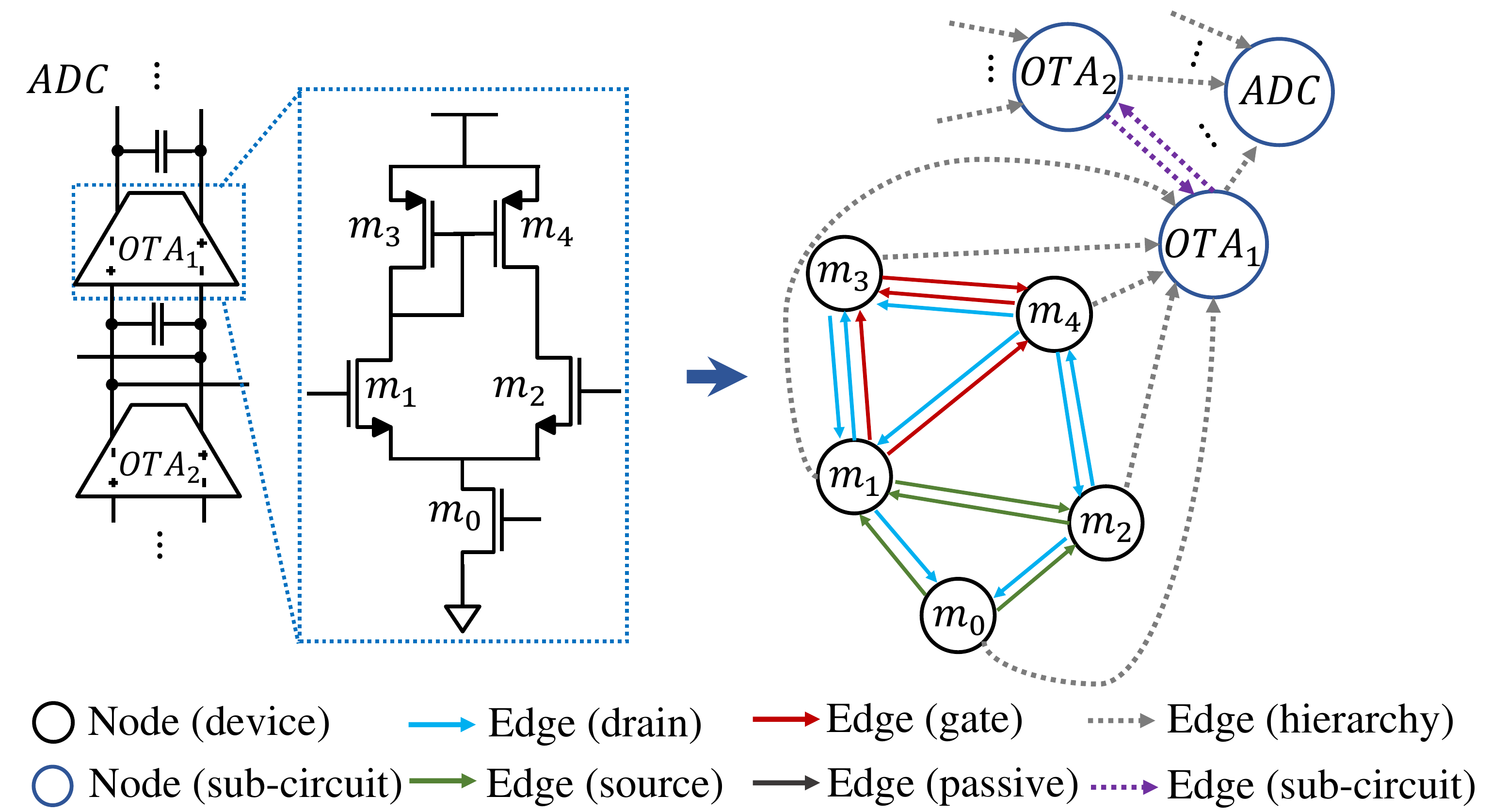}
    \caption{An example of the graph representation for an AMS circuit.}
    \label{fig:graph}
\end{figure}

Figure~\ref{fig:graph} illustrates an example of the proposed graph representation.
The example includes three hierarchy levels.
Nodes $m_0, \dots, m_4$ represent the transistors composing $OTA_1$.
One hierarchy edge directs from every transistor node to the $OTA_1$ sub-circuit node denoting the hierarchy.
On a higher level, the interconnections between $OTA_1$ and $OTA_2$ are represented with the two-pin pair model similar to the leaf nodes.
The proposed graph representation maintains a homogeneous structure between different hierarchy levels and enables our learning model to apply to device-level and circuit-level prediction.

\subsection{Instance Feature Initialization}
\label{sec:algo:feature}

In TAG, each instance has two sets of features.
The first feature set includes the conventional features such as node type, sizing, and area. 
We use this set of features in the GNN operation and name it graph node features.
The other set is the text embeddings of instance names and instance types.
We pre-train a word embedding model and use the learned text embeddings to provide additional information to the graph node features.

\subsubsection{Device Parameter Features}

We initialize the graph's first feature set for device and sub-circuit nodes.
The first part of the node feature is a one-hot vector of node types.
In this work, the node types include regular NMOS, regular PMOS, thick gate NMOS, thick gate PMOS, resistor, capacitor, and sub-circuit.
The second part is an instance's width, height, and area.
For sub-circuits, we sum up the area of children instances to approximate the sub-circuit areas.
The widths and heights are then calculated, assuming the aspect ratio is 1.
The third part of the feature is the sizing parameters.
They define the device geometries and influence the circuit functionalities. 
Table~\ref{tab:features} lists the parameters included.
All the parameters are normalized.
We average their children's instances to obtain the sizing feature fields for sub-circuit nodes.

\begin{table}[tb]
\centering
\caption{List of Instance Parameters.}
\resizebox{0.3 \textwidth}{!}{
\begin{tabular}{|c|c|c|}
\hline\hline
\textbf{Type}         & \textbf{Feature}                      & \textbf{Definition}                   \\ \hline\hline
\multirow{3}{*}{Transistors}     & L                 & Gate poly length                    \\ \cline{2-3}    
                                        & NF                 & Number of fingers                   \\ \cline{2-3}  
                                         & NFIN                 & Number of fins                   \\ \hline    
\multirow{2}{*}{Resistors}               & L         & Length of resistors            \\ \cline{2-3} 
                                  & W                & Width of resistors      \\ \hline
Capacitors                        & M                & Multipliers    \\ \hline\hline
\end{tabular}
}
\label{tab:features}
\end{table}

\begin{figure}[tb]
    \centering
    \subfloat[]{\includegraphics[width=0.23\textwidth]{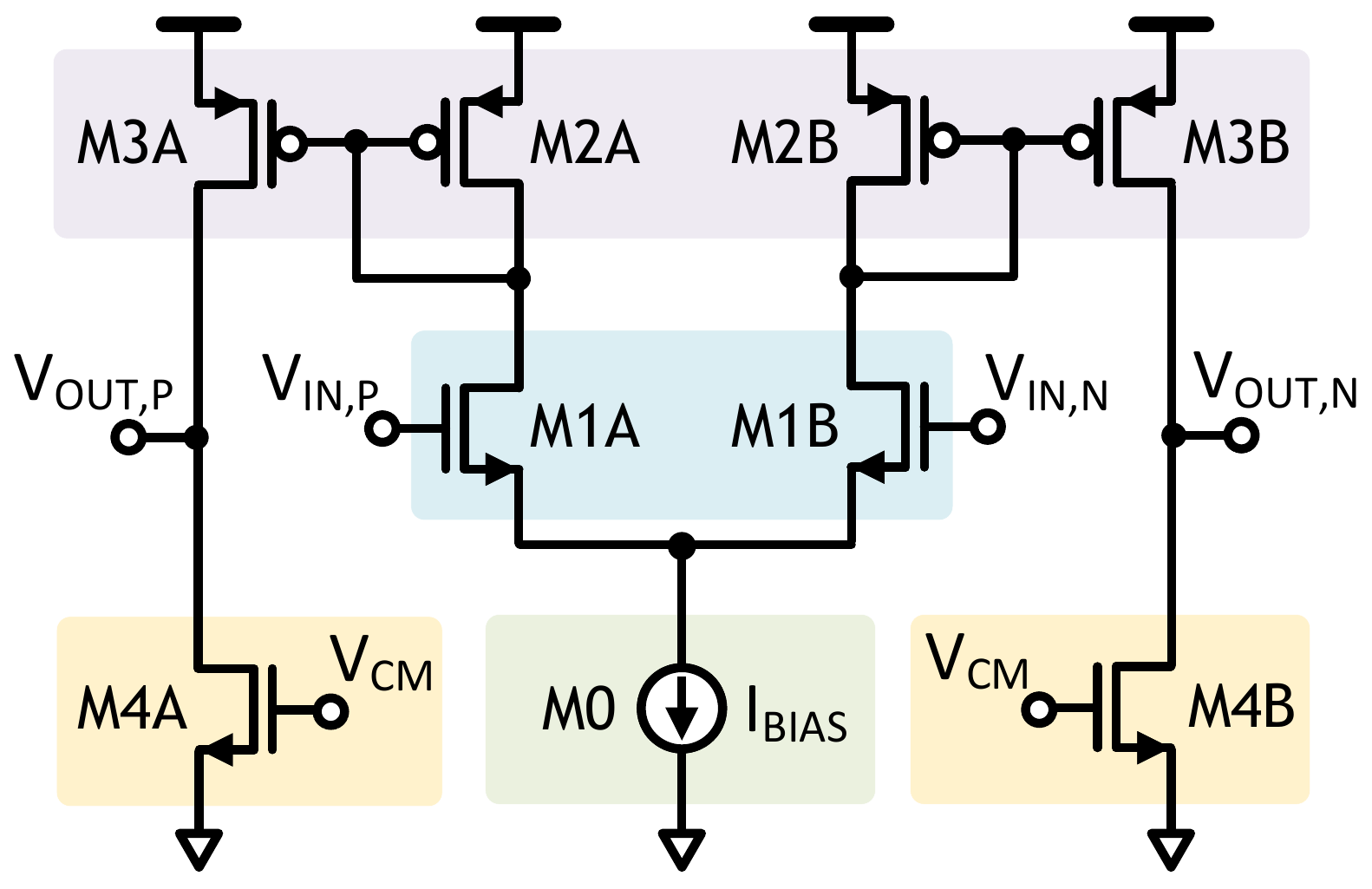}}
    \hfill
    \subfloat[]{\includegraphics[width=0.09\textwidth]{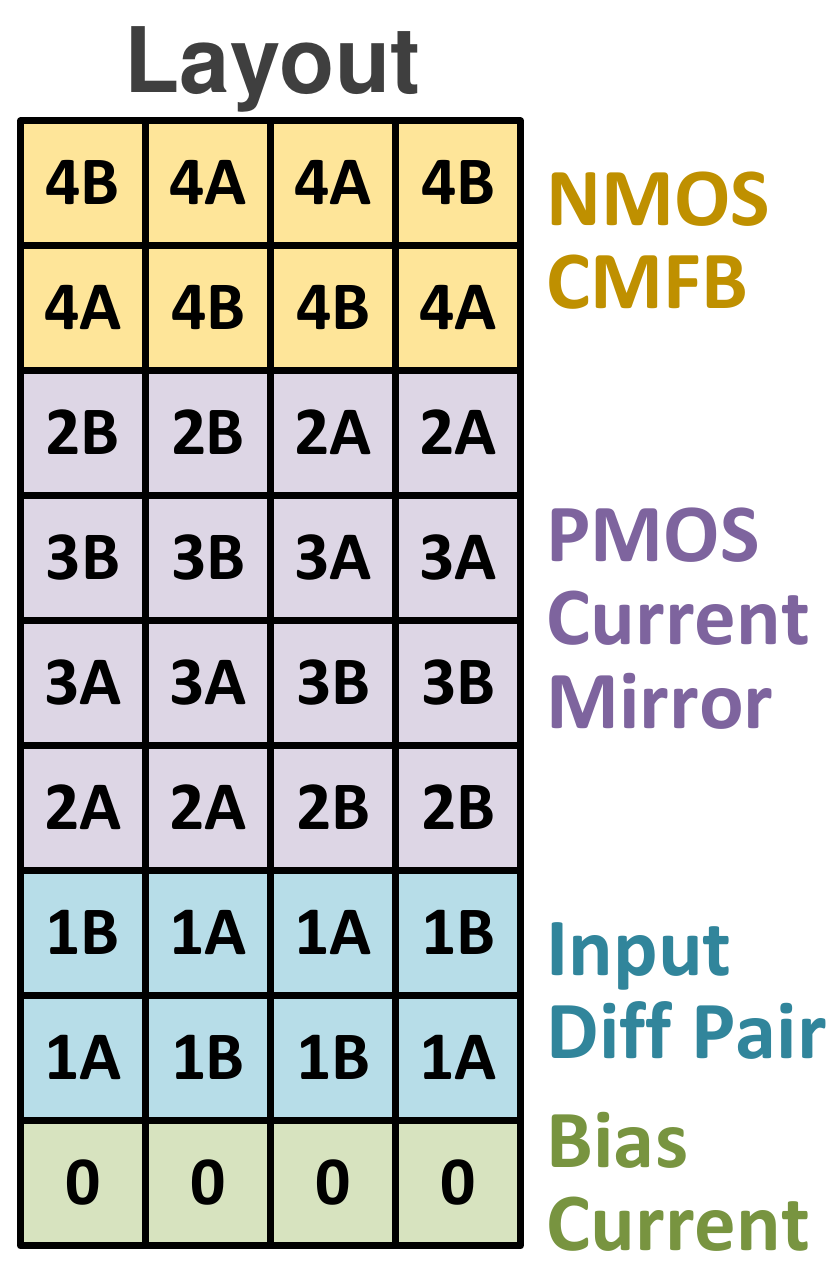}}
    \caption{An OTA design with symmetric structure. (a)~The schematic. (b)~Manual layout abstraction.}
    \label{fig:ota}
\end{figure}

\begin{figure}[tb]
    \centering
    \subfloat[]{\includegraphics[width=0.23\textwidth]{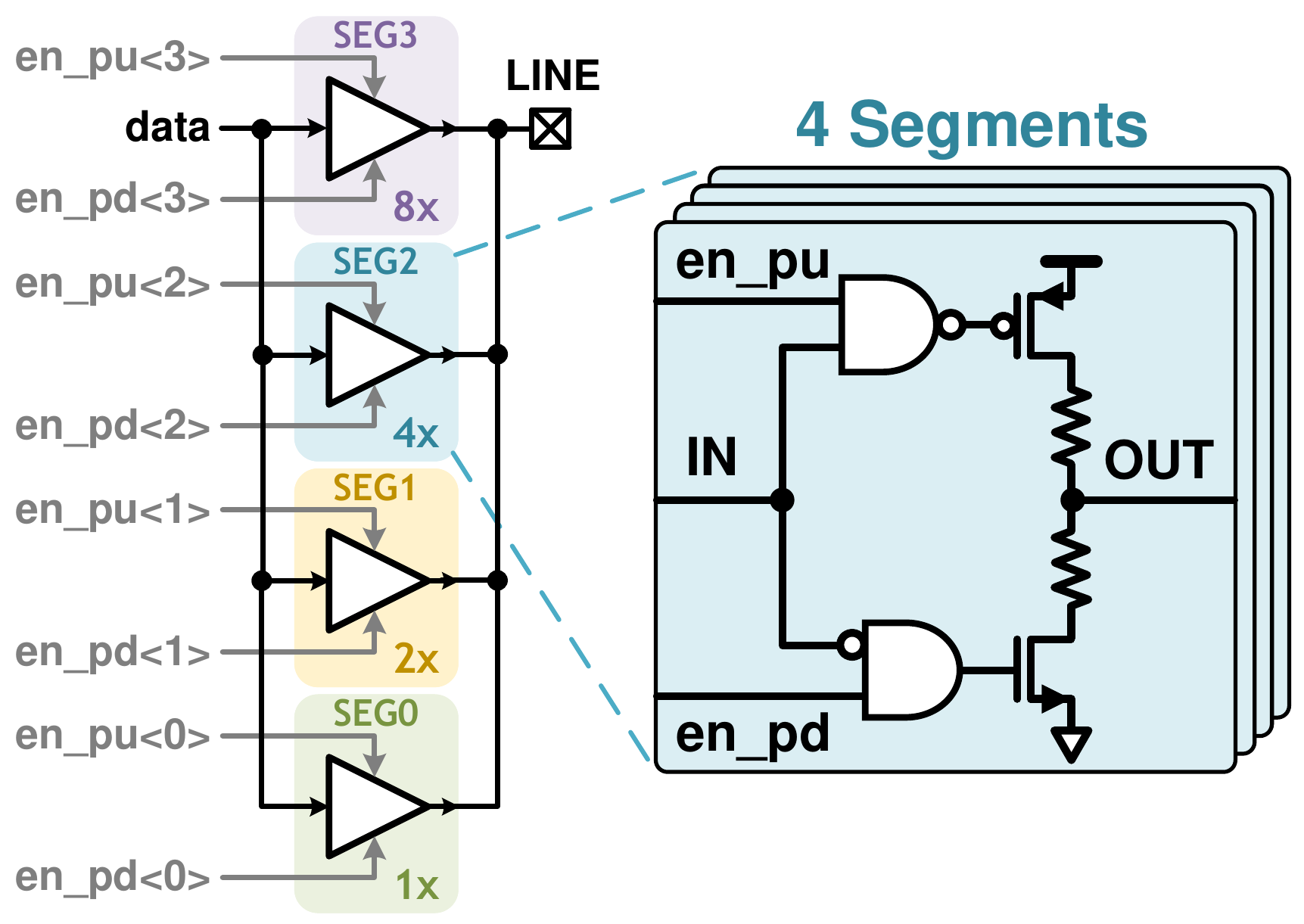}}
    \hfill
    \subfloat[]{\includegraphics[width=0.09\textwidth]{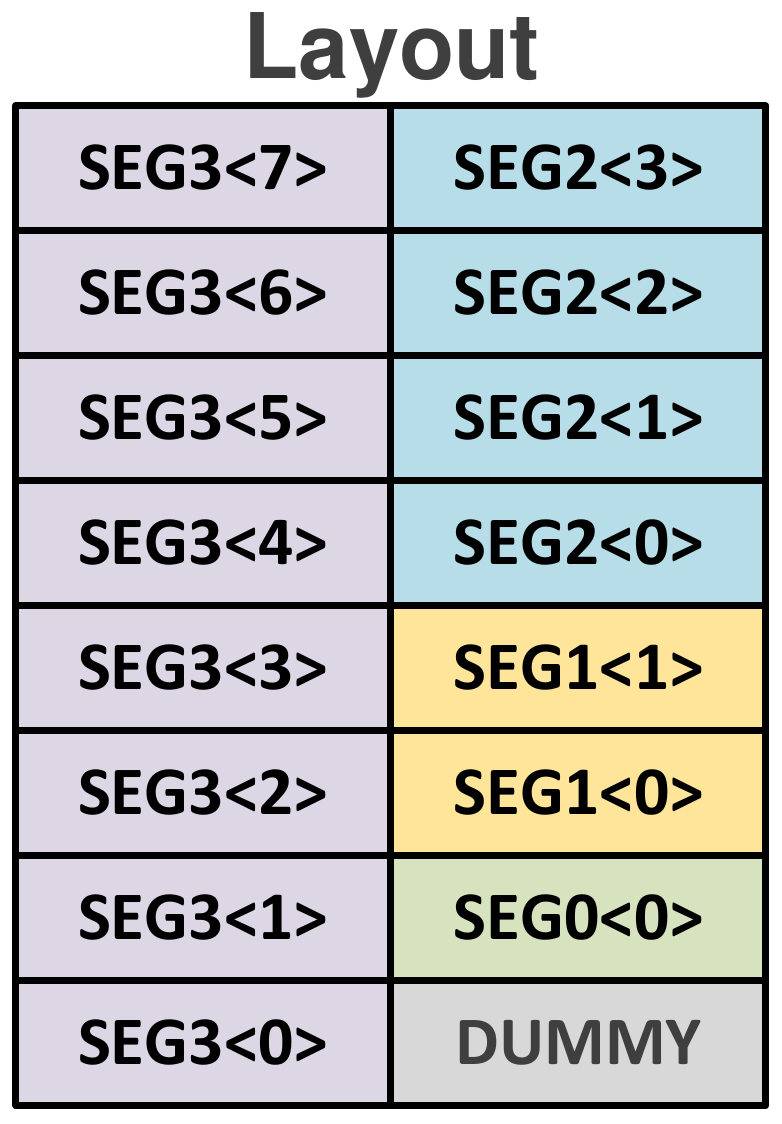}}
    \caption{A line driver design with array structure. (a)~The schematic. (b)~Manual layout abstraction.}
    \label{fig:driver}
\end{figure}

\subsubsection{Texts Features}

In addition to the graph node feature, we also incorporate the word embedding of the instance name and device/sub-circuit type name in our framework.

Circuit netlists describe instances using an associated instance name and a device type, where the use of this information has been usually overlooked so far.
In typical AMS circuit netlists, an instance is associated with an instance name from the circuit netlist and an instance type.

The instance names empirically incorporate the purpose and the position of the instances.
Intuitively, the designers select the names to help them understand the circuit design, e.g., \texttt{NMOS0}, \texttt{INV0}, and \texttt{NDIFF}.
Although not always deliberately planned, the name of an instance usually provides prior knowledge from design expertise.
We observe in real-world designs that there is a correlation between the naming similarity and placement proximity.
The instance names can be utilized as supplementary knowledge in circuit representation learning.

We also find that the instance names are beneficial for offsetting some of the limitations of ConvGNNs.
The instance names help distinguish locally isomorphic structures while retaining an inductive basis across different circuits.
Figure~\ref{fig:ota} shows an example of an operational transconductance amplifier~(OTA).
Its schematic has a highly symmetric structure.
As a result, typical ConvGNNs usually are challenged in distinguishing the \texttt{A} branch nodes from the \texttt{B} branch in the embedding space, such as \texttt{M4A} and \texttt{M4B}.
On the other hand, by examining the instance names, a human can quickly identify the circuit structures.
Because those names contain prior knowledge of pair-wise symmetry relations and instance positions in the circuit, the naming convention plays an even more critical role in mixed-signal designs where parallel structures are common.
Figure~\ref{fig:driver} shows a line driver design.
The driver consists of unit-sized driver segments configured for binary-weighted digital control of drive strength.
The only difference between the segments is the connection to different control signals (e.g., \texttt{en\_pu<3:0>} and \texttt{en\_pd<3:0>}), which are also coming from control circuits with similar structures.
Such differences are challenging for ConvGNNs to learn, while humans can easily understand by observing the instance naming.

Device/sub-circuit type names are also often neglected in existing AMS circuit learning schemes.
A common approach (e.g., ~\cite{PD_ML_DAC20_Ren}) is to group the devices into several groups, such as PMOS, NMOS, and capacitors.
However, such a paradigm omits the detailed difference between device types, such as low and high threshold voltage transistors.
On the other hand, the work~\cite{MAGICAL_CICC21_Chen} separates every device type with a one-hot encoding scheme.
However, it adds additional complexity for the network to learn the behavior of every single device.
Besides, there are few considerations of sub-circuit names in circuit learning models.
Although circuit type identification techniques exist~\cite{Analog_DATE20_Kunal}, a principal method to vectorize sub-circuit type for learning tasks is yet to be explored.
On the other hand, the device types and circuit types are usually well described in their naming for circuit designers to understand.
Similar device types also have parasitics that are correlated.

In TAG, we consider the texts in the circuit representation learning tasks.
In~\cite{PLACE_ISPD21_Lu}, names of module hierarchies are encoded with trie~(suffix graph) to assist placement.
The modules with common ancestors in the hierarchy tree have similar hierarchical encoding.
The similarity of hierarchies benefits the ML-assisted placement to find better clustering of the modules.
However, the trie-based encoding method does not leverage the semantic information of the texts and is hard to be extended to new designs.
A new paradigm is proposed to vectorize instances and type names using word embeddings.

We adopt the fastText framework~\cite{ML_ARXIV16_Bojanowski} to vectorize the texts.
It extracts the subword information to enable learning from words having similar subwords~(e.g., \texttt{nch\_ulvt\_mac} and \texttt{nch\_lvt\_mac}).
It uses a hashing function to store the dictionary that allows producing word embedding for unseen words.
We extract the sentences from the netlists of 1490 industrial AMS circuit designs.
The following words are treated as one sentence:
(1) the current circuit name and the device/sub-circuit type names of its children instances,
and (2) the instance name, its device/sub-circuit type names, and the net names connecting to this instance.
We then combine the extracted sentences with the first 1 billion bytes of English Wikipedia corpus~\cite{enwik9}.
We train the word embedding model using the fastText framework with a word embedding dimension of 64, a context window size of 10, and a maximum length of character N-gram of 15.

\subsection{Instance Representation Embedding Network}
\label{sec:algo:embed}

\begin{figure}[t]
    \centering
    \includegraphics[width=0.48\textwidth]{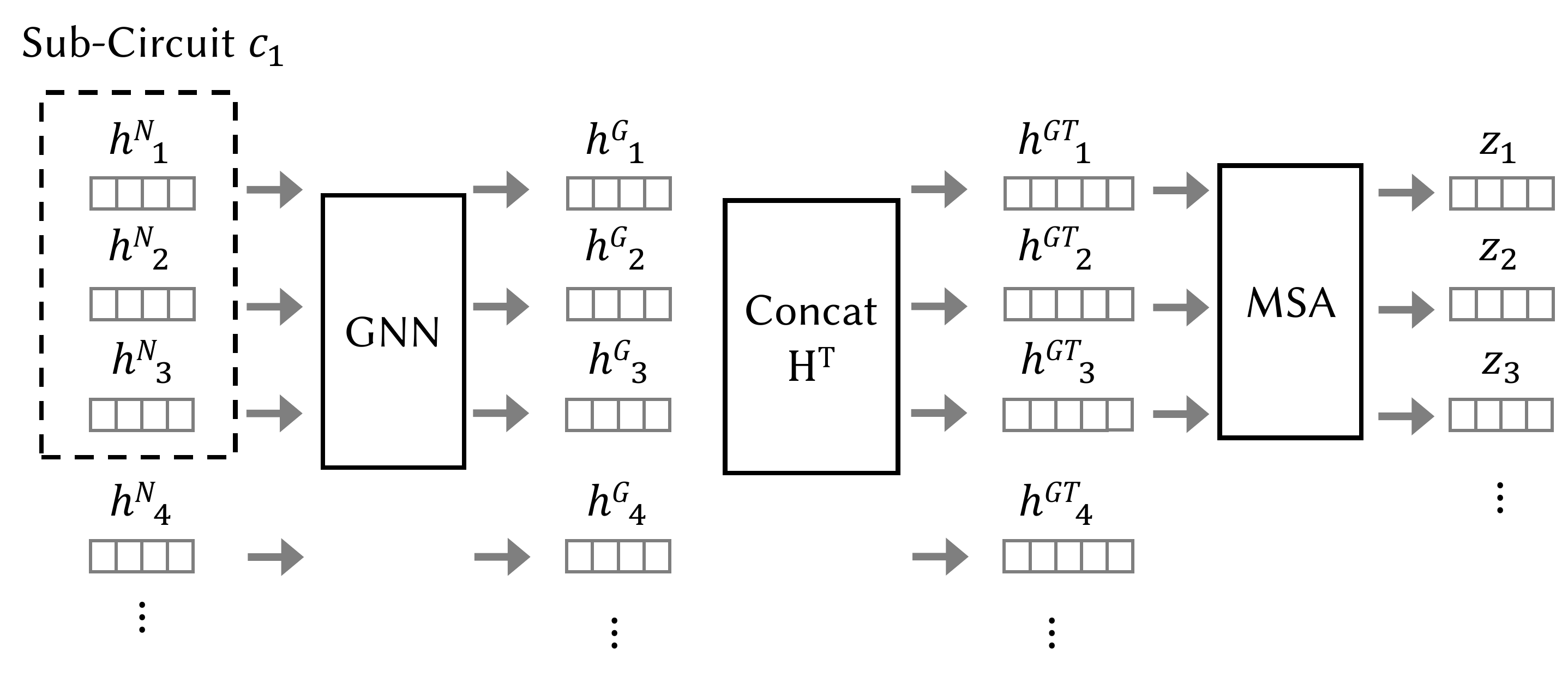}
    \caption{Illustration of the proposed instance representation embedding network.}
    \label{fig:flow}
\end{figure}

The embedding network in TAG contains two stages: the GNN and the multi-head MSA stage.
Algorithm~\ref{alg:embed}  sketches the procedures.
We first compute the graph embeddings $\mathbf{H}_G$ using a GNN model~(Line~1).
Then we concatenate the graph embedding $\mathbf{H}^G$ with the instance text embedding $\mathbf{H}^T$ and apply linear transform on it to form a combined embedding $\mathbf{H}^{GT}$~(Line~2).
We then treat the combined embeddings $\mathbf{H}^{GT}$ from the same sub-circuit as a collection~(Line~6).
This collection of embedding vectors is sent to an MSA layer~(Line~7).
The MSA layer enables all the instances in their sub-circuit to be considered together, adding a global context to the final instance embeddings.
Figure~\ref{fig:flow} shows an illustration of the proposed TAG network.
After the MSA layer, the TAG embedding vectors $\mathbf{Z}$ are then fed to the distance prediction network or the other downstream tasks.

\begin{algorithm}[t]
  \caption{Instance Embedding Algorithm in TAG}
  \label{alg:embed}
  \begin{algorithmic}[1]
    \Require A heterogeneous hierarchical graph representation $G = (V, E)$, node features $\mathbf{H}^N$, text embedding features $\mathbf{H}^T, \forall i \in V$ and a set of sub-circuits $C$.
    \Ensure The instance embeddings $\mathbf{Z}$.
    \State GNN forward operation $\mathbf{H}^G = \mathrm{GNN}(G, \mathbf{H}^N)$.
    \State Concatenate with text embeddings $\mathbf{H}^{GT} = \mathrm{concat}(\mathbf{H}^G, \mathbf{H}^T)$
    \State Transform to embedding dimension $\mathbf{H}^{GT} = \mathbf{W} \mathbf{H}^{GT}$
    \State Initialize an empty matrix $\mathbf{Z} = \mathrm{zeros}(|V|, d)$
    \ForEach{Sub-circuit $C_i \in C$}
        \State Extract the instances embeddings $\mathbf{H}_{C_i} = \{\mathbf{H}^{GT}_{k}, \forall k \in C_i\}$
        \State MSA forward operation $\mathbf{Z}_{C_i} = \mathrm{MSA}(\mathbf{H}_{C_i})$
    \EndFor
    \Return $\mathbf{Z}$
  \end{algorithmic}
\end{algorithm}

The GNN network in TAG has two convolution layers.
The first layer uses different linear transform matrices for different edge types to distinguish different edge connections.
The convolution operation on this layer is shown in Equation~\eqref{eq:etypegnn}
\begin{equation}
\label{eq:etypegnn}
\mathbf{h}^{l+1}_i = \mathrm{ReLU}\left( \mathbf{W}^l_{self} \mathbf{h}^l_i + \mathrm{mean}\left(\sum_{j \in \mathcal{N}_i} \mathbf{W}^{l}_{e_{ij}} \mathbf{h}_j^{l}\right)\right),
\end{equation}
where $\mathbf{h}^{(l)}_i$ is the $l^\mathrm{th}$ layer output for node $n_i$, $\mathcal{N}_i$ indicates the neighbors of node $n_i$, $\mathbf{W}^l_{self}$ is the weight for transform on the node $n_i$ itself and $\mathbf{W}^l_{e_{ij}}$ is the weight for edge type $e_{ij}$.
The second layer is a graph isomorphism network~(GIN) layer~\cite{ML_ICLR19_Xu}.
It is provably as powerful as the Weisfeiler-Lehman graph isomorphism test and is shown in Equation~\eqref{eq:gin}.
\begin{equation}
    \label{eq:gin}
    \mathbf{h}_i^{(l+1)} = \mathbf{W} \left( (1 + \epsilon) \mathbf{h}_i^{l} +
        \sum_{j\in\mathcal{N}_i}\left\{\mathbf{h}_j^{l}
        \right\}\right),
\end{equation}
where $\mathbf{W}$ is a weight matrix, and we use $\epsilon = 0$ in the experiments.
In our implementation, we set the hidden layer dimension and the output embedding dimension in the GNN to be 64 and 32, respectively.
We also experiment with the variants of GNN with popular ConvGNN layers and change the number of layers.
The impact of GNN architecture choice is relatively minor compared to our other proposed techniques.

The graph embedding and pre-trained text embedding of the same node are concatenated together and sent to an MSA layer. 
 We embed instances within a sub-circuit as an unordered sequence, on which we apply the self-attention mechanism.
Self-attention~(SA)~\cite{ML_NIPS17_Vaswani} is a popular building block for machine learning on sequences. 
Equation~\eqref{eq:sa} shows its computation equations,
\begin{equation} \label{eq:sa}
    \begin{aligned} 
    [\mathbf{q}, \mathbf{k}, \mathbf{v}] &= \mathbf{z} \mathbf{U},  &&\mathbf{U} \in \mathbb{R}^{D \times 3 D_{h}}, \\
    A         &= \mathrm{softmax}(q k^T / \sqrt{D_{h}}),  && A \in \mathbb{R}^{N \times N}, \\
    \mathrm{SA}(\mathbf{z}) &= A\mathbf{v}, && 
    \end{aligned}
\end{equation}
where $q$, $k$, and $v$ are query, key, and value matrices of each embedding, $N$ is the sequence length, $A$ is the attention of each query-key pair, and $\mathrm{SA}(\mathbf{z})$ is the final embedding of each node based on the attention over the value matrices of other nodes.
The SA mechanism can be applied to an arbitrary input sequence length.
MSA  extends the SA mechanism to run $k$ SA operations, called ``head'', in parallel.
The MSA operation is shown in Equation~\eqref{eq:msa}.
\begin{equation}
\label{eq:msa}
\begin{aligned}
   \mathrm{MSA}(\mathbf{z}) &= [\mathrm{SA}_1(\mathbf{z}); \mathrm{SA}_2(\mathbf{z}); ...; \mathrm{SA}_k(\mathbf{z})] \mathbf{U},   & \mathbf{U} \in \mathbb{R}^{k\cdot D_h \times D}.
    \end{aligned}
\end{equation}
We use $k = 4$, $D = 64$, and $D_h = 16$ in the experiments.

\subsection{Layout Instance Distance Prediction Loss}
\label{sec:algo:dist}

After obtaining the embeddings, we iterate through all instance pairs in the same sub-circuit and predict their relative distances in the manual layout implementation.
This learning task allows the embedding network to extract the knowledge from human layout implementation without additional manual labeling.

An ad-hoc approach to predict distance based on two embedding vectors is to concatenate them and feed-forward using a fully connected network~(FC) as shown in Equation~\eqref{eq:cat}.
\begin{equation}
    \label{eq:cat}
    y_{ij} = \mathrm{FC}( [\mathbf{z}_i; \mathbf{z}_j])
\end{equation}

\begin{figure}[t]
    \centering
    \includegraphics[width=0.28\textwidth]{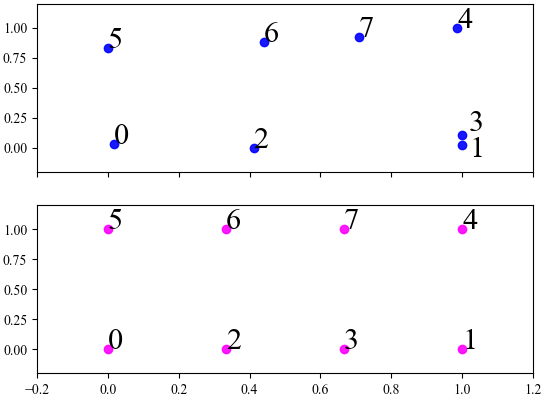}
    \caption{Example of the spatial embeddings and their corresponding layout locations. Above: The two dimensional  principal component analysis of the embeddings. Bottom: The layout locations.}
    \label{fig:pca}
\end{figure}

We also propose a more direct approach for predicting the relative distance in TAG.
We assume an instance embedding vector space exists where the distance in this space is proportional to the expected placement distance in manual layout.
Equation~\eqref{eq:norm} shows our proposed method.
\begin{equation}
    \label{eq:norm}
    \begin{aligned}
    \mathrm{NORM}(i, j, C, \mathbf{H}) &= \frac{ \norm{\mathbf{h}_i - \mathbf{h}_j}} { \max_{k, l}(\norm{\mathbf{h}_k -\mathbf{h}_l })}, & \forall k,l \in C, k\neq l,
    \end{aligned}
\end{equation}
where $\mathbf{H}$ denotes the embedding space, $\mathbf{h}_i$ indicates the embedding of instance $I_i$ and $C$ is a collection of instances, which is the sub-circuits in our scenario.
The denominator term finds the maximum distance between the instance pair in this sub-circuit, approximating the sub-circuit diameters.
Intuitively, we measure the relative layout distance of two instances by computing their distance normalized by the sub-circuit diameters.
Figure~\ref{fig:pca} shows an illustration of our learned $\mathbf{H}$ and the corresponding instance relative positions in a sub-circuit.
In the implementation, we use the $\mathrm{LogSumExp}(x_i,\dots, x_n) = \log(\exp(x_i)+\dots+\exp(x_n))$ function to smooth and approximate the $\max(\cdot)$ function for more robust and efficient training.
TAG adopts the scheme and adds a layer normalization step and an FC network before the distance norm computation, as shown in Equation~\eqref{eq:norm_FC}.
\begin{equation}
    \label{eq:norm_FC}
    \mathrm{DIST}(i,j,C,\mathbf{Z}) = \mathrm{NORM}(i, j , C, \mathrm{FC}(\mathrm{LayerNorm}(\mathbf{Z}))),
\end{equation}
where $Z$  is the embedding after MSA layer.
We choose to add an additional fully connected (FC) layer before the NORM layer because we empirically find doing so leads to better transferability to the downstream tasks. 
 $L_2$ norm and FC networks with 1 hidden layer of dimension 128 in our experiments are used.
The mean squared error loss is used to train the model.
\section{Experimental Results}
\label{sec:result}

We implement the framework in Python with the PyTorch library.
All models are trained on a single NVIDIA Tesla V100 GPU with 32GB memory.
All models are trained with an ADAM optimizer.

The proposed method is evaluated on a dataset of 447 industrial AMS circuits in sub-10nm technology.
The size of the circuits ranges from 20 to 2000 instances.
We exclude the sub-circuits under four instances to avoid the results being dominated by na\"{i}ve cases and sample 20 instances from one sub-circuit for large sub-circuits.
To extract the placement coordinates of each device in the layout view, we use the StarRC extraction tool.

The circuits in the dataset are randomly shuffled and split into training, validation, and test sets with 60\%, 20\%, and 20\% allocation, respectively.
Because different circuits sometimes share common sub-circuits, to avoid data leakage, we exclude all the sub-circuits that appear in the training set when doing the validation and testing.
We report the test set results in the experiments at the epoch with the lowest validation loss.

The training time takes about 10 hours on the dataset.
The inference time for each circuit takes an average of 0.09 seconds and a max of 0.8 seconds.
Most of the inference time is spent reading the files instead of model inference.
As the training is a one-time job, the runtime for TAG is considered negligible in usual applications.

To evaluate the solution quality, we adopt two sets of statistical measurements.
For regression tasks, we use R-squared~($R^2$), Mean Absolute Error~(MAE), and symmetric Mean Absolute Percentage Error~(sMAPE) as the metrics.
For binary classification tasks, we adopt accuracy~(ACC), true positive rate~(TPR), false positive rate~(FPR),  positive predictive value~(PPV),  and $F_1$-score over the valid pairs.
Higher $R^2$, ACC, TPR, PPV, and $F_1$ scores are better, while lower MAE, sMAPE, and FPR scores are better.

To evaluate the effectiveness of our learned circuit representation in applications, we obtained the source codes of the AncstrGNN~\cite{MAGICAL_DAC21_Chen} and Paragraph~\cite{PD_ML_DAC20_Ren} from the authors.
We train these frameworks on our dataset and compare our circuit representation in the case studies.

In the rest of this section, we evaluate our text embedding quality and the circuit representation learning scheme and conduct two case studies for using our model in other two learning tasks: detecting matching in layouts and predicting the wirelength.

\subsection{Circuit Text Embedding}

We first evaluate the quality of the word embedding model.
The word embedding ideally shall provide a meaningful similarity measurement between the instances.
We verify this property by directly applying the distance norm method (Equation~\eqref{eq:norm}) on the word embedding to predict the relative placement distances.
As shown in Table~\ref{tab:direct_distance}, our model yields an $R^2$ of 0.205.
This result is already better than the vanilla GNN-only approach, even without layout data or additional trainable parameters.
It shows the proposed text embedding contains valuable information for learning.
In comparison, the same model pre-trained with natural language corpus (Common Crawl) alone can only result in a $R^2$ of -0.783.
The improvement of our word embedding model shows the benefits of training the word embedding model with sentences extracted from netlists.

\begin{table}[tb]
\centering
\caption{Comparisons  of  $R^2$, MAE and MAPE for directly predicting instance relative distance with text embedding distance norm.}
\resizebox{0.4 \textwidth}{!}{
\begin{tabular}{|c|c|c|c|}
\hline\hline
Method         &   $R^2$      & MAE                &sMAPE      \\ \hline\hline
NLP Pre-trained Model~\cite{ML_LREC18_Mikolov}             & -0.783                 & 0.291              & 0.565        \\ \hline    
Our model          & \textbf{ 0.205}         & \textbf{0.186}    & \textbf{0.452}             \\ \hline \hline
\end{tabular}
}
\label{tab:direct_distance}
\end{table}

\begin{figure}[tb]
    \centering
   \includegraphics[width=0.46\textwidth]{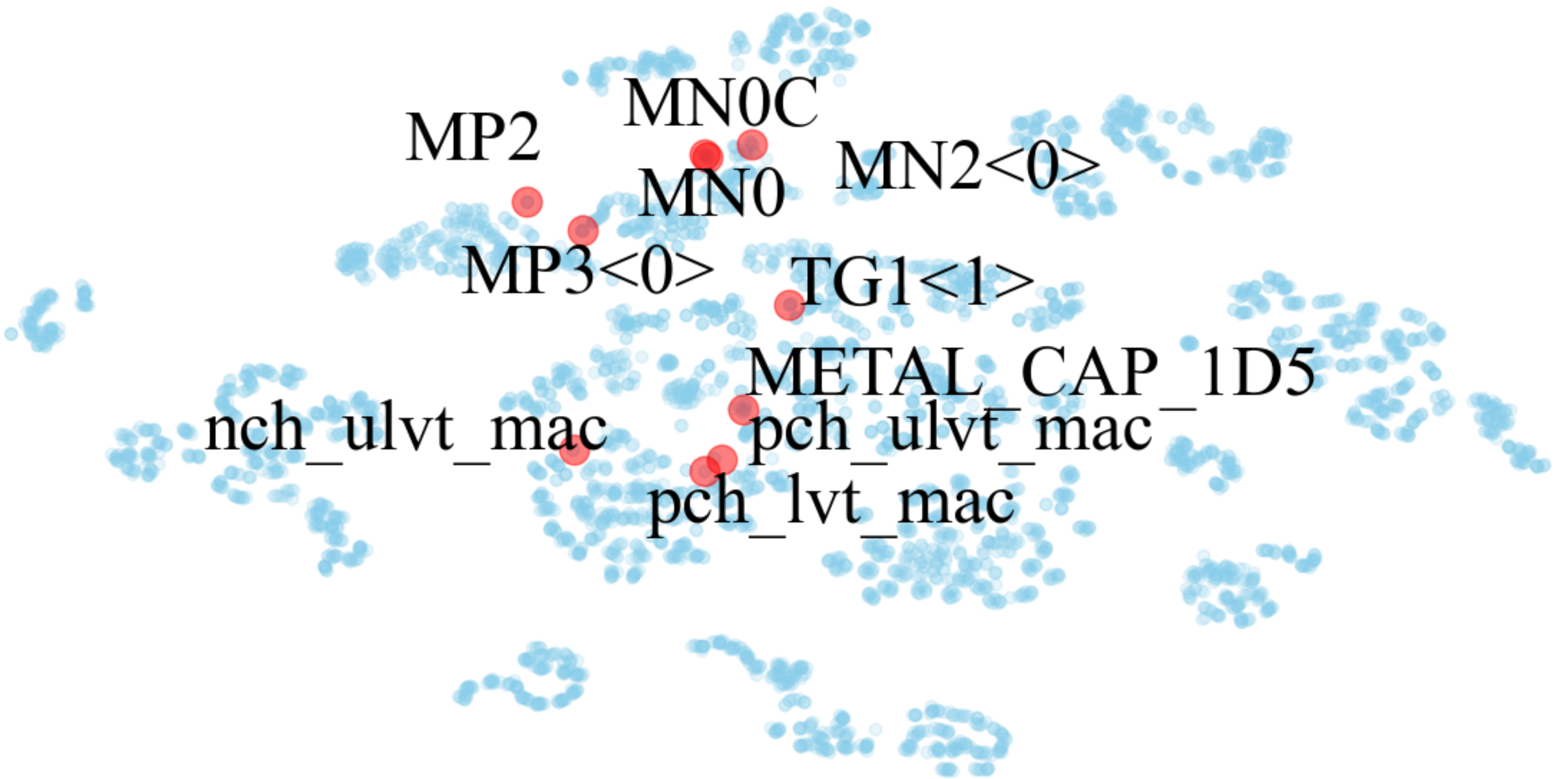}
    \caption{t-SNE plots of embeddings of the word embeddings. }
    \label{fig:tsne}
\end{figure}

To investigate the meaning of the embeddings, we visualize the high-dimensional embedding vector with the t-SNE algorithm~\cite{ML_JMLR08_Maaten}.
Figure~\ref{fig:tsne} shows the t-SNE plots for the words in 30 circuits.
It is observed that two PMOS~(\texttt{pch\_ulvt\_mac} and \texttt{pch\_lvt\_mac}) device types are close and well separated from the NMOS device type~(\texttt{nch\_ulvt\_mac}).
The t-SNE plot illustration shows that our word embedding model captures the similarities in texts and provides a new dimension of information and the conventional graph representation.

Within our benchmark circuits, many instance names are, in fact, not explicitly named, e.g., \texttt{M\_I1} and \texttt{XI0}.
However, from our experience, the important instances are usually well named.
Our experimental results demonstrate the overall effectiveness of the text embeddings even with the existence of arbitrary naming conventions in some netlists.

\subsection{Instance Relative Distance Prediction}

We evaluate our model accuracy on instance relative distance prediction.
We predict the distances between all the instance pairs within the same sub-circuit.
The distance is normalized to the sub-circuit placement bounding box.
Table~\ref{tab:distance} shows the regression results compared with different variants of the models and training methods.
The model types are denoted before the dash in the method names.
``T'', ``A'' and ``G'' indicate the model contains text embedding, self-attention layers, and a graph neural network, respectively.
The prediction method is labeled after the dash.
``CAT'' denotes concatenating the instance pair embeddings and predicting the distance with an FC network~(Equation~\eqref{eq:cat}).
``NORM'' indicates to use our proposed distance measuring method~(Equation~\eqref{eq:norm_FC}).
For example, TG-CAT denotes the model using GNN and text embedding and predicts the distance with the concatenation method. 
T-NORM denotes using only text embedding without graph with distance norm method.

Our proposed TAG framework outperforms the other methods in all metrics and achieves $R^2$ of 0.64.
Meanwhile, the conventional GNN-only structure can hardly produce meaningful predictions above 0 $R^2$.
Because layout instance distances are intrinsically noisy due to the manual implementation, our proposed model demonstrates a strong capability to learn the circuit representation.
From the ablation study with different model variants, we also observe that each component of our proposed framework benefits the learning task.
These observations demonstrate the effectiveness of our proposed techniques.

We also compare with the AncstrGNN~\cite{MAGICAL_DAC21_Chen}.
``AncstrGNN-A'' denotes using the AncstrGNN.
AncstrGNN uses a Gated-GNN layer to generate node embeddings based on a contrastive loss between nodes. 
We adopt the "CAT" approach for relative distance prediction to predict pair-wise distance from pair embeddings. 
The sub-circuit embeddings are obtained by mean aggregating their children's embeddings.
``AncstrGNN-B'', on the other hand, uses the AncstrGNN network architecture but with our proposed hierarchical graph representation.
In both cases, AncstrGNN can not learn the relative distance effectively. The results are similar to our G-only model. 

Note that the ``G-NORM'' and ``AG-NORM'' options both fail in training.
The reason is rooted in the graph formulation and the ConvGNN mechanism.
As discussed in Section~\ref{sec:prelim}, the local isomorphic structure will make nodes indistinguishable.
As a result, the distance between two node embeddings might be close or equal to zero.
When we use the distance norm method, the backward gradient in such a case will be very large and cause the training process to diverge.
This observation also shows the importance of text embedding.

\begin{table}[tb]
\centering
\caption{Comparisons  of  $R^2$, MAE and sMAPE for instance relative distance prediction.}
\resizebox{0.28 \textwidth}{!}{
\begin{tabular}{|c|c|c|c|}
\hline\hline
Method         &   $R^2$      & MAE                &sMAPE      \\ \hline\hline
AncstrGNN~\cite{MAGICAL_DAC21_Chen}-A       & -0.091                  &    0.225        & 0.508       \\ \hline    
AncstrGNN~\cite{MAGICAL_DAC21_Chen}-B       & 0.068                  &    0.191        & 0.502       \\ \hline    
G-CAT              & 0.075                 & 0.134              & 0.502        \\ \hline    
T-CAT              & 0.194                 & 0.187             & 0.489        \\ \hline    
TA-CAT             & 0.452                 & 0.164             & 0.453        \\ \hline    
TG-CAT             & 0.335                 & 0.177              & 0.542        \\ \hline    
AG-CAT             & 0.367                 & 0.184              & 0.508        \\ \hline    
TAG-CAT             & 0.585                 & 0.134              & 0.404        \\ \hline   
G-NORM              & FAIL                 & FAIL              &  FAIL       \\ \hline    
T-NORM              & 0.321                 & 0.177             & 0.458        \\ \hline    
TA-NORM             & 0.530                 & 0.140              & 0.409        \\ \hline    
TG-NORM             & 0.470                 & 0.154              & 0.442        \\ \hline    
AG-NORM             & FAIL                 & FAIL              & FAIL        \\ \hline    
TAG-NORM          & \textbf{ 0.640}         & \textbf{0.122}    & \textbf{0.364}             \\ \hline \hline
\end{tabular}
}
\label{tab:distance}
\end{table}

\subsection{Application Case Study 1: Layout Matching Prediction}

The effectiveness of our learned circuit representation is evaluated for predicting the matching patterns in the layout.
We use the same dataset in the pre-training process.
Instructed by the designers, the labels of matched instances are extracted based on layout coordinates.
The matching conditions include interdigitation pattern, common-centroid pattern, and symmetry pattern.
The methods are evaluated to detect those matching pairs.
This task is similar to the symmetry constraint detection problem.

We randomly selected 10\% of the entire dataset from the training set to train the models.
Another 10\% of circuits from the validation set in the previous stage are used to validate the task.
The entire test set (20\% of circuits) is used for testing.
The task is treated as a binary classification task by concatenating two instance embeddings and forwarding with a two-layer FC network and is trained with the cross-entropy loss.

We compare the proposed pre-trained circuit representation with training from scratch and the state-of-the-art symmetry detection framework AncstrGNN~\cite{MAGICAL_DAC21_Chen}.
Table~\ref{tab:matching} shows comparisons of evaluation metrics.
``AncstrGNN-A-CAT'' concatenates the pre-trained AncstrGNN embedding and trains an FC network to do binary classification.
``AncstrGNN-A-COS'' uses the cosine similarity criteria to predict the symmetry constraint as proposed in the original paper.
``AncstrGNN-B-CAT'' and ``AncstrGNN-B-COS'', on the other hand, are using our proposed hierarchical graph representation. 
``G trained from scratch'' and `TAG trained from scratch'' train the network without pre-trained weights.
``TAG'' is with our proposed pre-trained embeddings.
The embedding network weights in this configuration are fixed in the training so that its results measure the generality of our pre-trained embeddings.
Our proposed method outperforms training from scratch and AncstrGNN with an $F_1$ score of 0.842.
This observation shows the effectiveness of applying the proposed pre-trained circuit representation in other tasks.

\begin{table}[tb]
\centering
\caption{Comparisons  of ACC, TPR, FPR, PPV and $F_1$ for layout matching prediction.}
\resizebox{0.48 \textwidth}{!}{
\begin{tabular}{|c|c|c|c|c|c|}
\hline\hline
Method                                    & ACC         & TPR               & FPR             & PPV             & $F_1$      \\ \hline\hline
AncstrGNN~\cite{MAGICAL_DAC21_Chen}-A-CAT     & 0.677       &  0.802            & 0.434            & 0.621           & 0.706    \\ \hline    
AncstrGNN~\cite{MAGICAL_DAC21_Chen}-A-COS     & 0.805        &    0.724         & \textbf{0.086}          & \textbf{0.919}            & 0.810\\ \hline    
AncstrGNN~\cite{MAGICAL_DAC21_Chen}-B-CAT     & 0.750       &  0.701            & 0.203            & 0.765           & 0.731    \\ \hline    
AncstrGNN~\cite{MAGICAL_DAC21_Chen}-B-COS     & 0.720        &    0.740         & 0.305           & 0.738            & 0.739\\ \hline    
G Trained from scratch                            & 0.731        &    0.706         & 0.246           & 0.730            & 0.718    \\ \hline    
TAG Trained from scratch                          & 0.730        &    0.666         &0.208   & 0.751            & 0.706    \\ \hline    
TAG Pre-trained                                & \textbf{0.833} & \textbf{0.915} & 0.244           & 0.780   & \textbf{0.842}    \\ \hline \hline
\end{tabular}
}
\label{tab:matching}
\end{table}

\subsection{Application Case Study 2: Wirelength Estimation}

Another case study is to estimate the net HPWL.
We also use the same dataset in the pre-training process.
Like the instance distance prediction task, we normalize HPWL with respect to the sub-circuit layout bounding box to allow inductive learning.
The 10\%/10\%/20\% data splitting is used for training, validation, and test sets similar to the matching prediction task.

We use a self-attention layer with four heads and mean aggregation on the instance embeddings and predict the HPWL using a two-layer FC network, as shown in Equation~\eqref{eq:hpwl}.
\begin{equation}
    WL = \mathrm{FC}\left(\mathrm{mean}\left(\mathrm{MSA}\left(\mathbf{Z}_N\right)\right)\right),
    \label{eq:hpwl}
\end{equation}
where $\mathbf{Z}_N$ denotes a collection of instance embeddings connected by net $N$.

Table~\ref{tab:hpwl} shows the comparisons of HPWL prediction results.
``AncstrGNN-A'' denotes using the AncstrGNN embedding with the original flatten graph, while ``AncstrGNN-B'' uses the TAG configurations.
The pre-trained TAG model achieves the best result in all evaluation metrics.
The observation in this case study aligns with the results from the matching prediction task that TAG outperforms the baselines.
It is observed that there is a performance gap between ``AncstrGNN-A'' and ``AncstrGNN-B''.
We believe that adding hierarchy knowledge with our proposed graph representation benefits the learning task.

\begin{table}[tb]
\centering
\caption{Comparisons  of $R^2$, MAE and sMAPE for relative HPWL prediction.}
\resizebox{0.4 \textwidth}{!}{
\begin{tabular}{|c|c|c|c|}
\hline\hline
Method                                    & $R^2$                   & MAE              &sMAPE      \\ \hline\hline
AncstrGNN~\cite{MAGICAL_DAC21_Chen}-A       & -0.177                  &    0.146        & 0.550       \\ \hline    
AncstrGNN~\cite{MAGICAL_DAC21_Chen}-B       & 0.203                  &    0.198        & 0.520       \\ \hline    
G Trained from scratch                            & 0.027                  &    0.219        & 0.566       \\ \hline    
TAG Trained from scratch                          & 0.153                  &    0.212        & 0.543       \\ \hline    
TAG Pre-trained                  & \textbf{0.570}         &  \textbf{0.139} & \textbf{0.469}             \\ \hline \hline
\end{tabular}
}
\label{tab:hpwl}
\end{table}

\subsection{Application Case Study 3: Net Parasitic Capacitance Prediction}

We also evaluate the effectiveness of our pre-trained embeddings in the net parasitic capacitance prediction task.
This task uses a dataset of 385 industrial AMS circuits in sub-10nm technology.
Based on the recommendation from the designers, we use 17 designs in the dataset as the testing set and the rest as the training set.
We verify that there is no overlap between this testing set and the training set in the model pre-training.

The TAG embeddings are integrated with the state-of-the-art parasitics prediction algorithm, ParaGraph~\cite{PD_ML_DAC20_Ren}.
The TAG embedding vectors for all the instances are first generated with a pre-trained TAG model.
Then we augment the ParaGraph input features with these TAG embeddings.
Five models are trained at one time with different maximum prediction values ($max_v$) of 0.5fF, 1fF, 10fF, 100fF and 1pF. 
The final prediction is obtained with the ensemble modeling technique as suggested in the original paper.

Table~\ref{tab:para_acc} shows the comparisons of the prediction accuracy.
With the augmented TAG embedding, the net parasitic capacitance prediction achieves significant improvement in accuracy for the 0.5fF and 1pF models.
It also produces similar accuracy for the 1fF, 10fF, and 100fF models.
We believe that it is because the pre-trained TAG embedding incorporates valuable spatial information of the circuits.
The additional spatial information allows the model to make more accurate predictions considering layout effects.
Table~\ref{tab:para_error} shows the corresponding errors on simulated performance.
With the more accurate net capacitance predictions, the TAG embedding helps to reduce the mean performance error from 18.6\% to 11.8\%.
The results demonstrate the effectiveness of our proposed TAG model.

\begin{table}[tb]
\centering
\caption{Comparisons  of $R^2$ and MAE of different $max_v$  for the net parasitic capacitance prediction task.}
\resizebox{0.36 \textwidth}{!}{
\begin{tabular}{|c|c|c|c|}
\hline\hline
 Metrics & $max_v$ & ParaGraph~\cite{PD_ML_DAC20_Ren} & TAG \\ \hline \hline
 \multirow{5}{*}{$R^2$} & 0.5fF & 0.495& \textbf{0.678} \\ \cline{2-4}
 & 1fF &  0.830 & \textbf{0.876}\\ \cline{2-4}
  & 10fF & 0.854 & \textbf{0.856} \\ \cline{2-4}
  & 100fF & \textbf{0.872} & 0.870 \\ \cline{2-4}
  & 1pF & 0.308 &  \textbf{0.411}\\ \hline
  \multirow{5}{*}{MAE} & 0.5fF & $2.71e-17$ & $\bm{2.25e-17}$ \\ \cline{2-4}
 & 1fF & $5.14e-17$ & $\bm{3.41e-17}$ \\ \cline{2-4}
  & 10fF & $2.01e-16$ & $\bm{1.83e-16}$ \\ \cline{2-4}
  & 100fF & $\bm{3.23e-16}$& $3.39e-16$ \\ \cline{2-4}
  & 1pF & $9.93e-16$ &  $\bm{9.07e-16}$\\ \cline{2-4} \hline
\end{tabular}
}
\label{tab:para_acc}
\end{table}

\begin{table}[tb]
\centering
\caption{Comparisons  of mean and geometric mean of the errors in simulated performance with predicted net parasitic capacitance.}
\resizebox{0.36 \textwidth}{!}{
\begin{tabular}{|c|c|c|}
\hline\hline
Method                                    & Mean                   & Geometric Mean                  \\ \hline\hline
ParaGraph~\cite{PD_ML_DAC20_Ren}       & 18.6\%                  &    4.97\%            \\ \hline    
TAG                                   & \textbf{11.8\%}         &  \textbf{4.17\%}             \\ \hline \hline
\end{tabular}
}
\label{tab:para_error}
\end{table}


\section{Conclusion}
\label{sec:concl}

This paper has presented TAG, a new paradigm and framework to learn and pre-train circuit instance representations.
Using relative layout distance for the training target, TAG embeds the high-level knowledge into the representation by fitting the spatial information.
It leverages the netlists and introduces sub-circuit-wise MSA to assist the training.
A comprehensive algorithm set has been presented, including feature extraction, network architecture, and learning algorithm.
Experimental results have demonstrated the efficiency and effectiveness of TAG on learning spatial knowledge and the ability to transfer the learned embeddings to other learning tasks.
\section*{Acknowledgement}

This work is supported in part by NVIDIA Corporation, the DARPA IDEA program, and the NSF under Grant No. 1704758 and 2112665.
The authors would like to thank Xuyang Jin from The University of Texas at Austin and Yanqing Zhang from NVIDIA Corporation for helpful comments and discussions.

\balance

{
\scriptsize
\small
 \footnotesize
\bibliographystyle{IEEEtran}
\bibliography{./ref/Top_sim, ./ref/MAGICAL, ./ref/analog, ./ref/PD, ./ref/analog_place.bib, ./ref/ML.bib, ./ref/general.bib, ./ref/cktml.bib, ./ref/other.bib}
}

\end{document}